\documentstyle[12pt]{article} 
\newlength{\extraspace}
\newlength{\extraspaces}
\newcommand{\be}{\begin{equation}
\addtolength{\abovedisplayskip}{\extraspaces}
\addtolength{\belowdisplayskip}{\extraspaces}
\addtolength{\abovedisplayshortskip}{\extraspace}
\addtolength{\belowdisplayshortskip}{\extraspace}}
\newcommand{\ee}{\end{equation}}

\newcommand{\ba}{\begin{eqnarray}
\addtolength{\abovedisplayskip}{\extraspaces}
\addtolength{\belowdisplayskip}{\extraspaces}
\addtolength{\abovedisplayshortskip}{\extraspace}
\addtolength{\belowdisplayshortskip}{\extraspace}}
\newcommand{\ea}{\end{eqnarray}}

\newcommand{\nonu}{\nonumber \\[.5mm]}
\newcommand{\A}{&\!\!\!}

\newcommand{\il}{\lambda_{{}_{{}_{\!\!\!\!\scriptstyle{i}}}}}

\newcommand{\bl}{\lambda_{{}_{{}_{\!\!\!\!\scriptstyle{0}}}}}

\newcommand{\blz}{\lambda_{{}_{{}_{\!\!\!\!\scriptstyle{0}}}}}

\newcommand{\al}{\lambda_{{}_{{}_{\!\!\!\!\scriptstyle{a}}}}}

\newcommand{\newsection}[1]{
\vspace{7mm}
\pagebreak[3]
\addtocounter{section}{1}
\setcounter{subsection}{0}
\setcounter{footnote}{0}
\begin{center}
{\large {\bf \thesection. #1}}
\end{center}
\nopagebreak
\medskip
\nopagebreak
\hspace{3mm}}

\begin{document}

\begin{center}
{{\bf Effect of signature change in {\rm NGR} }}
\end{center}
\centerline{ F. I. Mikhail\footnote{\rm {Late Professor of
Mathematics, Ain Shams University.}}, M. I. Wanas$\,{}^{\dagger}$
,and G. G. L. Nashed$\,{}^{\ddagger}$}

\bigskip
\centerline{${}^\dagger$\ {\it Astronomy  Department, Faculty of
Science, Cairo University-Giza  , Egypt}}
\bigskip

\bigskip

\centerline{${}^{\ddagger}$\ {\it Mathematics Department, Faculty
of Science, Ain Shams University-Cairo, Egypt }}

\bigskip
 \centerline{${}^\dagger$\ e-mail:Wanas@frcu.eun.eg}

\hspace{2cm}
\\
\\
\\
\\
\\
\\
\\
\\

The field equations of the new general relativity (NGR) have been
applied to an absolute parallelism space having  three unknown
functions of the radial coordinate. The field equations have been
solved using two different schemes. In the first scheme, we used
the conventional procedure used in orthodox general relativity. In
the second scheme, we examined the effect of signature change. The
latter scheme gives a solution which is different from the
Schwarzschild one. In both methods  we find  solution of the field
equations under the same constraint  imposed on the parameters of
the theory.  We also calculated the energy associated with the
solutions in the two cases using the superpotential method. We
found that the energetic content of one of the solutions is
different from that of the other.  A comparison between the two
solutions obtained in the present work and a third one obtained by
Hayashi and Shirafuji (1979) shows that the change of the
signature may give rise to new physics.
\newpage
\begin{center}
\newsection{\bf Introduction}
\end{center}

  It is generally accepted that a successful theory of gravity
should  be a metric one.  In applications of such a theory, it is
necessary to attribute a property called signature to the metric
representing space-time (cf. \cite{MT}). The signature usually
chosen, in most applications, is the Lorentz signature
(corresponding to an indefinite metric). This signature guarantees
a desirable result that any gravity theory of this class will have
correct special relativistic limits. However, it appears that
there is a need for another signature different from the Lorentz
one, giving rise to a positive definite metric, especially when
dealing with some problems connected  to quantum  cosmology
\cite{HH}. It is well known that a complete consistent theory that
combines  quantum mechanics and gravity is still beyond the reach
of researchers. However, one important features of such  a theory,
 if it exists, is that it should incorporate Feynman's proposal to
formulate quantum theory in terms of a sum over histories. Severe
technical problems arise when one tries to apply Feynman's
proposal in this context. The only way to overcome these problems
is that one must add up waves for particle histories not in the
real time but in the imaginary  time \cite{H}. This necessitates
the use of a positive-definite metric and not an indefinite one.
So, it is of interest to examine the effect of the signature of
the metric on solution of different geometric theories of gravity.

Some authors started to examine the effect of signature change on
the solutions of the field equations of general relativity \cite
{W,ES,TM,E,JH,TE}. The attempts of those authors can be classified
in two different classes:\\  {\it The  first class,} The main
assumption in this class is that the signature of the metric is
changed, from a positive-definite metric to an indefinite one, at
a very early time in the history of the universe. This means that
the change is represented as a jump process across a certain
hypersurface (cf.\cite{ES}). \\ {\it The second class,} This class
depends on a different assumption. That is, the space-time has
4-dimensions with a positive-definite metric, but measurements are
carried out in (3+1)-dimensions with an indefinite metric giving
rise to Lorentz signature \cite{W}. Consequently, Lorentz
signature is to be imposed on the metric just before matching the
results of the theory with measurements. In other words, the field
equations of the theory are to be solved using a positive-definite
metric and then, after obtaining the solution, Lorentz signature
is to be imposed on the metric.

The effects of signature change were examined, in the previous
mentioned attempts, in the case of general relativity. It is of
interest to examine these effects on the solutions of other metric
field theories different from general relativity. Geometric field
theories of this class are those theories written in absolute
parallelism {\rm spaces} (cf.\cite{MW,M,KS}). Recently, more
attention is paid to such theories since they include
non-vanishing torsion which is connected to Dirac field \cite{RH}
and to string theory \cite{RH8}. One of those theories is the New
General Relativity (NGR) constructed by Hayashi and
  Shirafuji \cite{KS}. The basic geometry for this theory is the
absolute parallelism (AP) geometry, in which a metric can always
be defined.

The aim of the present work is to study the effects of signature
change on the spherically symmetric solutions of the NGR and to
calculate the energy associated with each solution, using the
superpotential method of Mikhail et al. \cite{MWHL}
 is calculated. In section 2 we briefly review the field
equations of NGR and the most general AP-structure having
spherical symmetry, used for applications. To explore the effect
of signature change on the solution, we are going, in section 3,
to solve the field equations for the spherically symmetric
structure, in two different cases. {\it Case I:} by imposing
Lorentz signature , on the metric, from the beginning, i.e.,
before solving the field equations; {\it Case II:}  we solve the
field equation; using a positive-definite metric then we impose
Lorentz signature to the solution obtained. In section 4 the
energy associated with each solution is calculated using the
superpotential method. We compare and discuss the results of
section 3 and 4 in section 5.

%\newpage
\newsection{NGR Field Equations and Geometric Structure}
%\newsection{Basic Lagrangian}
In 1979 Hayashi and Shirafuji constructed a theory which they
called "The New General Relativity", (NGR). They have used an AP-
space for its formulation, with the field variables being the 16
tetrad components $({\il}^{\mu})$\footnote {Latin indices
$(i,j,k,\cdots)$ designate the vector number, which runs from 0 to
 3, while Greek indices $(\mu,\nu,\rho, \cdots)$ designate the world-vector components
running from 0 to 3. Latin indices $(a,b,c,\cdots)$, and Greek
indices $(\alpha, \beta,\gamma,\cdots)$, run from 1 to 3.}.
Assuming  invariance under: \vspace{.3cm}
\\ a) the group of general coordinate transformations, and
\vspace{.3cm} \\ b) the group of global Lorentz transformations.
\vspace{.3cm} \\
 They wrote the general gravitational Lagrangian density quadratic in the torsion
  tensor as\footnote {
Throughout this paper we use the relativistic units, $c=G=1$. The
Einstein constant ${\kappa}$ is then $8 \pi $. We will denote the
symmetric part by the parentheses ( \ ), for example, $A_{(\mu \nu)}
 \stackrel {\rm def.}{=}(1/2)( A_{\mu \nu}+A_{\nu \mu})$ and the
antisymmetric part by the brackets [\ ], $A_{[\mu \nu]} \stackrel
{\rm def.}{=} (1/2)( A_{\mu \nu}-A_{\nu \mu})$.} \be {\cal L}_G=
\sqrt{-g} \left[{R \over 2 \kappa} +d_1(t^{\mu \nu \lambda} t_{\mu
\nu \lambda})+d_2(C^{\mu}
C_{\mu})+d_3(a^{\mu}a_{\mu})-d_4(C^{\mu}a_{\mu})\right], \ee where
$d_1$, $d_2$, $d_3$ and $d_4$ are dimensionless parameters of the
theory \footnote{ The dimensionless parameters $c_i$ of Ref.[12]
are here denoted by $d_i$ for convenience.}, and \ba
 t_{\mu \nu \lambda} \A \stackrel{\rm def.}{=} \A  \Lambda_{(\mu \nu) \lambda} -{1
\over 3}
 g_{\lambda (\mu}C_\nu)+{1 \over 3}g_{\mu \nu}C_\lambda, \nonu
{\Lambda^\lambda}_{\mu \nu} \A \stackrel{\rm def.}{=} \A
{\Gamma^\lambda}_{\mu \nu}-{\Gamma^\lambda}_{\nu \mu} =
{\il}^\lambda ({\il}_{\mu,\nu}-{\il}_{\nu,\mu}),\qquad {\rm
(Torsion \quad tensor),} \nonu
{\Gamma^\lambda}_{\mu \nu} \A \stackrel{\rm def.}{=} \A
{\il}^\lambda {\il}_{\mu,\nu},\qquad {\rm (Nonsymmetric \quad
connection),} \nonu
 C_\mu \A \stackrel{\rm def.}{=} \A {\Lambda^\lambda}_{ \mu  \lambda},\qquad
 {\rm (Basic \quad vector),} \nonu
a_\mu \A \stackrel{\rm def.}{=} \A {1 \over 6} \epsilon_{\mu \nu
\lambda \rho} \Lambda^{ \nu \lambda \rho},\qquad {\rm (Axial \quad
vector),} \nonu
 \epsilon_{\mu \nu \lambda \rho} \A \stackrel{\rm def.}{=} \A \sqrt{-g}
\delta_{\mu \nu \lambda \rho}, \qquad {\rm (Completely \quad
antisymmetric),} \nonu
g_{\mu \nu} \A \stackrel{\rm def.}{=} \A {\il}_{\mu} {\il}_{\nu},
\qquad {\rm (Metric \quad tensor),}
 \ea
 $\delta_{\mu \nu \lambda \rho}$ is the
Levi-Civita antisymmetric tensor and $R$ is the Ricci scalar.

 By applying the variational principle to the Lagrangian
(1), they obtained the field equation:
\be
I^{\mu \nu}= {\kappa}T^{\mu \nu},
 \ee
 where,
\be I^{\mu \nu} \stackrel{\rm def.}{=} G^{\mu \nu}+2{\kappa}
\Biggl[{D}_\lambda F^{\mu \nu \lambda}- C_\lambda F^{\mu \nu
\lambda}+H^{\mu \nu} -{1 \over 2} g^{\mu \nu} L' \Biggr], \ee and
\be F^{\mu \nu \lambda} \stackrel{\rm def.}{=} {1 \over 2}
{\il}^\mu {\partial L_G \over
\partial {\il}_{\nu,\lambda}} =-F^{\mu \lambda \nu}, \ee
\be
H^{\mu \nu} \stackrel{\rm def.}{=} \Lambda^{\rho \sigma \mu}
 {F_{\rho \sigma}}^\nu - {1 \over 2} \Lambda^{\nu \rho \sigma}
{F^\mu}_{\rho \sigma}=H^{\nu \mu},
\ee
\be
 L' \stackrel{\rm def.}{=} \left[d_1(t^{\mu \nu \lambda} t_{\mu \nu \lambda})+d_2(C^{\mu}
C_{\mu})+d_3(a^{\mu}a_{\mu})-d_4(C^{\mu}a_{\mu})\right], \ee
\be
T^{\mu \nu} \stackrel{\rm def.}{=} {1 \over \sqrt{-g}} {\delta
{\cal L}_M \over \delta {\il}_\nu} {\il}^\mu. \ee
Here ${\cal
L}_M$ denotes the Lagrangian density of material fields, of which
 the energy-momentum tensor $T^{\mu \nu}$ is nonsymmetric in general.

In the case of ${\it static}$, ${\it spherically}$ ${\it
symmetric}$ space-time, with tetrad vector fields having a
diagonal form, the field equations (4) have been exactly solved
\cite{KS,KS1}. The exact solution  obtained is the same as that
obtained by assuming the invariance under parity operation
\cite{SN}.

In order to reproduce the correct Newtonian limit, the parameters
$d_1$ and $d_2$ should satisfy the condition \cite{KS} \be
d_1+4d_2+9d_1d_2=0, \ee which is called the Newtonian
approximation condition. This condition can be satisfied by
taking, \be d_1=-{1 \over 3(1-\epsilon)}, \quad d_2={1 \over
3(1-4\epsilon)}, \ee where  $\epsilon$ is a dimensionless
parameter. Comparison with Solar - System observations indicates
that $|\epsilon|$ must be very small.
%\newsection{Tetrad Space with Spherical Symmetry}

The structure of the AP-spaces with spherical symmetry has been
studied by  Robertson \cite{R}. The tetrad vectors defining
completely this structure in Cartesian coordinates can be written
as \ba {\blz}^0 \A = \A A(r),\nonu
{\blz}^\alpha \A = \A D(r) X^\alpha,\nonu
{\al}^0 \A = \A E(r) X^a \nonu
{\al}^\alpha \A = \A { \delta_a}^\alpha B(r)+F(r) X^a X^\alpha+
\epsilon_{a \alpha c} S(r) X^c,
 \ea
  where $A(r)$, $D(r)$, $E(r)$,
$F(r)$, $B(r)$ and $S(r)$ are functions of the radial coordinate
\\ $r=(X^a X^a)^{1/2}$. It  has been shown that \cite{R}
\vspace{.3cm}:
\\ 1) Improper rotation  is admitted if $S(r)$=0. \vspace{.3cm} \\
2) The functions $E(r)$ and $F(r)$ can be eliminated by mere
coordinate transformations, leaving the tetrad in the simple form

\be
 \left. \begin{array}{ll} {\blz}^0 = A(r),\vspace{.2cm}\\ {\blz}^\alpha =
D(r) X^\alpha, \vspace{.2cm}\\ {\al}^\alpha = {\delta_a}^\alpha
B(r).
\end{array} \right \}
\ee
 %\ba
 %{\blz}^0 \A=\A A(r), \nonu
%
%{\blz}^\alpha \A=\A D(r) X^\alpha,
%
%{\al}^\alpha \A=\A {\delta_a}^\alpha B(r).\nonu
 %\ea

It is of interest to point out that the tetrad vectors used
previously  to obtain an exact solution \cite{KS} is a special
case of the tetrad (12) when the function $D(r)=0$. Thus one may
expect to obtain more general solutions when the tetrad (12) is
applied to the field equations (4). The tetrad (12) can be written
 in the form:
\be
\left({\il}^\mu \right)= \left( \matrix{ A(r) & D(r) r & 0 & 0
\vspace{3mm} \cr 0 & B(r) \sin\theta \cos\phi & \displaystyle{B(r)
\over r}\cos\theta \cos\phi
 & -\displaystyle{B(r) \sin\phi \over r \sin\theta} \vspace{3mm} \cr
0 & B(r) \sin\theta \sin\phi & \displaystyle{B(r) \over r}\cos\theta \sin\phi
 & \displaystyle{B(r) \cos\phi \over r \sin\theta} \vspace{3mm} \cr
0 & B(r) \cos\theta & -\displaystyle{B(r) \over r}\sin\theta  & 0
\cr } \right), \ee in  spherical polar coordinates
$(t,r,\theta,\phi)$. Consequently, the metric tensor of the
Riemannian space, associated with the AP-space (13) can be written
in  the form (using (2))
 \ba
 g_{0 0} \A
=\A {(B(r)^2+ D(r)^2 r^2) \over A(r)^2 B(r)^2}, \quad g_{0 1}=g_{1
0}=-{D(r) r \over A(r) B(r)^2}, \quad g_{1 1} = -{1 \over B(r)^2},
\nonu
g_{2 2} \A=\A -{ r^2 \over  B(r)^2}, \quad
g_{3 3}={ r^2 \sin\theta^2 \over  B(r)^2}.
\ea

\newpage
\newsection{Solution Of The Field Equation}

For later  convenience, we will redefine the functions A and D
such that $A\rightarrow c^\star A$ and $D\rightarrow {c_1}^\star
D$. This will not affect the geometric structure (13). Now
substituting into the field equations (4) using (13), we get the
following  set of differential equations ($c^\star$ and
${c_1}^\star$ are parameters)
 \ba \A \A
(-1+\kappa (d_1+4d_2)) c^{\star 2}A  \Biggl \{ B^2 \Biggl(
2\epsilon
  \left( {  {A'} \over A} \right)'+ 2(1-2\epsilon)
  \left( {  {B'} \over B} \right)'
+ {4 \over r} \left[ \epsilon
   {  A' \over A} +(1-2\epsilon)
   {  B' \over B} \right] -2\epsilon{A'B' \over AB} \nonu
\A \A -\epsilon \left( {  {A'} \over A} \right)^2-
(1-4\epsilon)\left( {  B' \over B} \right)^2 -3{ (1-\epsilon)
c_1^{\star 2} D^2\over B^2} \Biggr) +c_1^{\star 2} D^2 r
\Biggl[l(r)- r \Biggl( s(r)+b(r) \Biggr)  \Biggr] \Biggr \}=0,
 \ea
\ba \A \A (-1+\kappa (d_1+4d_2)) c^\star c_1^\star AD\Biggl \{ B^2
\Biggl( 2(1-2\epsilon)
   {  A' \over A} + 2{ B' \over B}
- r \Biggl(2(1-2\epsilon)
   {  A'B' \over A B} +\epsilon
   \left ({  A' \over A} \right)^2 + \left ({  B' \over B}
\right)^2 \nonu
\A \A+3{  (1-\epsilon)c_1^{\star 2} D^2 \over B^2} \Biggr) \Biggr)
-c_1^{\star 2} D^2
 r^2 \Biggl[ r \Biggl( s(r)+b(r) \Biggr)-l(r) \Biggr] \Biggr \}=0,
\ea

\ba \A \A (-1+\kappa (d_1+4d_2)) c^\star c_1^\star AD\Biggl \{ B^2
\left[ 2(1-2\epsilon)
   {  A' \over A} + 2(1-3\epsilon) { B' \over B}+
8\epsilon { D' \over D} \right] - B^2 r \Biggl( 2(1-3\epsilon)
   {  A'B' \over A B}  \nonu
\A \A +\epsilon
   \left ({  A' \over A} \right)^2
+ (1-6\epsilon) \left ({  B' \over B} \right)^2+2\epsilon{  A'D'
\over A D}+2\epsilon{B'' \over  B} +6\epsilon{  B'D' \over B
D}-2\epsilon{  D'' \over D} +3{  (1-\epsilon) c_1^{\star 2} D^2
\over B^2} \Biggr) \nonu
\A \A
 -c_1^{\star 2} D^2 r^2 \Biggl[ r \Biggl( s(r)+b(r) \Biggr)-l(r) \Biggr] \Biggr \}=0,
\ea

\ba \A \A (-1+\kappa (d_1+4d_2))  \Biggl \{ B^2 c_1^{\star 2}D^2 r
\left[2(2-3\epsilon)
  {  A' \over A} + 2(4-3\epsilon){ B' \over B}-2(1-\epsilon)
{ D' \over D} \right]
\nonu
\A \A-  B^2 c_1^{\star 2} D^2 r^2 \Biggl(2(2-3\epsilon) {A'B'
\over AB} + 2\epsilon \left( {  {A'} \over A}
\right)^2+(4-3\epsilon) \left( { B' \over B} \right)^2-2\epsilon
{A'D' \over AD}- 2(1-\epsilon) {B'D' \over BD}+\epsilon \left( {
D' \over D} \right)^2 \nonu
\A \A + 3{  (1-\epsilon)c_1^{\star 2}D^2 \over B^2} \Biggr)+B^4
\Biggl[ 2(1-2\epsilon) {A'B' \over AB} +\epsilon \left( {  {A'}
\over A} \right)^2+
 \left( {  B' \over B} \right)^2-{2 \over r} \left(
(1-2\epsilon){A' \over A} +{ B' \over B} \right) \nonu
\A \A + 3{  (1-\epsilon)c_1^{\star 2}D^2 \over B^2}
\Biggr]+c_1^{\star 4}D^4 r^3 \Biggl[ r \Biggl( s(r)+b(r)
\Biggr)-l(r) \Biggr] \Biggr \}=0, \ea

\ba \A \A (-1+\kappa (d_1+4d_2)) B^2 \Biggl \{ c_1^{\star 2} D^2
\Biggl[(1-2\epsilon)
  {  A'' \over A} -3(1-2\epsilon) {A'B' \over AB}
-(2-5\epsilon)\left( {  {A'} \over A} \right)^2
-5(1-\epsilon) \left( {  B' \over B} \right)^2 \nonu
\A \A +(3-8\epsilon) {A'D' \over AD}-2(1-2\epsilon){B'' \over B} +
(5-8\epsilon) {B'D' \over BD}-(1-2\epsilon){D'' \over D}-
(1-3\epsilon)\left( {D' \over D} \right)^2 \Biggr] \nonu
\A \A+{c_1^{\star 2}D^2 \over r} \Biggl( 4(1-2\epsilon){A' \over
A}+8(1-\epsilon){B' \over B}- 2(3-5\epsilon){D' \over D}
\Biggr)+{B^2 \over r^2} \Biggl((1-2\epsilon){A'' \over A}-
 2\epsilon{A'B' \over AB}-(2-5\epsilon) \left({  {A'} \over A} \right)^2 \nonu
\A \A - \left({  B' \over B} \right)^2 +{B'' \over B}+{1 \over r}
\left[(1-2\epsilon){A' \over A}+{B' \over B} \right]- 3{
(1-\epsilon)c_1^{\star 2}D^2 \over B^2} \Biggr) \Biggr \} =0, \ea
where,
 \ba
 \epsilon \A \stackrel {\rm def.}{=} \A { \kappa
(d_1+d_2) \over (-1+\kappa (d_1+4d_2)) }, \nonu
l(r) \A  \stackrel {\rm def.}{=} \A 2(1-\epsilon)\left[{A'\over A}
+3{B' \over B}- {D' \over D}  \right], \nonu
s(r) \A \stackrel {\rm def.}{=} \A \left[\epsilon\left( {  {A'}
\over A} \right)^2 +3(1-\epsilon)\left( {  B' \over B} \right)^2+
\epsilon\left( {D' \over D} \right)^2 \right],\nonu
b(r) \A \stackrel {\rm def.}{=} \A 2\left[(1-\epsilon) {A'B'\over
AB}- \epsilon {A'D' \over A D}- (1-\epsilon){B'D' \over B D}
\right], \ea and $A' \stackrel {\rm def.}{=} \displaystyle{dA
\over dr}$, \quad $B' \stackrel {\rm def.}{=} \displaystyle{dB
\over dr}$ and \quad $D' \stackrel {\rm def.}{=} \displaystyle{dD
\over dr}$.

We are going to find a general solution of the differential
equations (15)-(19) following the method  given by Mazunder and
Ray \cite{MR} for  two cases, i.e, the indefinite and
positive-definite
cases.\hspace{-1cm}\\
{\it Case I: indefinite metric} \\
\hspace{.5cm} In this case we take the value of $c^\star=
c_1^\star=\sqrt{-1}$.
 Using (15), (16)  (taking $\epsilon=0$, to simplify
the equations) we get \be (r B'-B)B A'+(B B''-B'^2)r A+A B B'=0,
\ee equation (21) can be integrated to give the function A, in
terms of the unknown function B, in the form \be
A=\displaystyle{k_1 \over \left(1-\displaystyle{r B' \over B}
\right)},
 \ee
  $k_1$ being a constant of integration.

From (15), using (22), we get after some rearrangements:
\ba \A \A
D^2 \left(2 B B'' r^3-5 B'^2 r^3-3 r B^2+8 r^2B B' \right)+ D
D'\left( 2 B B' r^3-2 r^2 B^2 \right) \nonu
\A \A + 2 r B^3 B''-3 r B'^2 B^2+4 B^3 B'=0.
 \ea
 Using the transformation,
\be
B=e^{ \alpha}, \quad z=ln~r, \quad D^2=\beta,
 \ee
 then (23) will give,
\be \left (2 \alpha_{z z}-3 {\alpha_z}^2+6 \alpha_z -3 \right)+ {
\beta_z \over \beta} \left(\alpha_z -1 \right)- {1 \over
\beta}e^{(2 \alpha - 2z)} \left (2 \alpha_{z z}- {\alpha_z}^2+2
\alpha_z \right)=0, \ee where $\alpha_z=\displaystyle{d\alpha
\over dz}$. Now the solution of (25) can be written in the form
\be \beta=D^2={e^{3(\alpha-z)} \over (1-\alpha_z)^2} \left
\{k_2+\alpha_z(\alpha_z-2)e^{(z-\alpha)} \right \}, \ee  $k_2$
being another constant of integration. Using (24), we rewrite
 (26) in the form
\be D^2=\displaystyle{1 \over \left(1-\displaystyle{r B' \over B}
\right)^2} {\left( B \over r\right)^3} \left \{ k_2+{r B' \over B}
\left( {r B' \over B} -2 \right) {r \over B} \right \}, \ee in
terms of the arbitrary function B. Hence we obtain the general
solution of the field equations (15)-(19) in the case
$c^\star={c_1}^\star=\sqrt{-1}=i$
 in terms of an arbitrary function B
for the special case $\epsilon=0$.

The line - element in this case is given by \be dS^2=-{(B^2-D^2
r^2) \over A^2 B^2} dt^2-2{D r \over A B^2} dr dt + {1 \over B^2}
\left( dr^2+r^2 d\Omega^2 \right), \ee where \be
d\Omega^2=d\theta^2+\sin\theta^2 d\phi^2. \ee
 Using the coordinate transformation \cite{W5}
\be dT=dt+{D r A \over B^2- D^2 r^2}dr, \quad \xi={(B^2-D^2 r^2)
\over A^2 B^2}, \quad and \quad R={ r \over B}, \ee then we can
eliminate the cross term in (28), and we finally get  \be
dS^2=-\xi dT^2 + {1 \over \xi} dR^2+R^2 d\Omega^2. \ee Using (22),
(27), the terms containing the derivatives of the arbitrary
function $B$ cancel out and we finally get, \be
\xi(R)=\left(1-{k_2 \over R} \right).
 \ee
 Taking $k_2=2m$, then (31)
will give rise to the Schwarzschild  metric. Thus in this case,
i.e. indefinite metric, we get nothing more than the Schwarzschild
field as a solution of the NGR field equations.

The tetrad (13) in {\it Case I} has been subject to two steps of
coordinate transformations from $(t, r, \theta, \phi)$ to  $(T, R,
\theta, \phi)$. We now apply a further transformation from $(T, R,
\theta, \phi)$ to the Cartesian coordinate $(T,X^a)$ with $a=1,2,$
and 3. The tetrad can be shown to have the form \ba {\blz}^0 \A=\A
\displaystyle {(1-r B') \over \left(1-\displaystyle{2m \over
r}\right)}, \nonu
{\blz}^\alpha \A=\A {x^\alpha\over r^{3/2}}
\sqrt{2m+B'^2r^3-2B'r^2} , \nonu
{\al}^0 \A = \A  \displaystyle{ix^\alpha \over r}
{\sqrt{2m+B'^2r^3-2r^2B'} \over (r-2m )} , \nonu
{\al}^\alpha \A = \A-i\displaystyle{(B' x^\alpha x^a -r) \over r}.
 \ea
 Tetrad (33) will be used latter in the calculations of the
 energy for the case of indefinite metric.\\
 \hspace{-1cm} {\it Case II: positive definite}\\

 As we mentioned  before, in this case we are going to postpone the
insertion of Lorentz signature till we find the solution of the
field equations (15)-(19). This can be achieved by taking
$c^\star=c_1^\star=1$. Following the same procedure as in {\it
Case I},
 (taking $\epsilon=0$), we get from (15) and (16)
\be
(rB'-B)BA'+(BB''-B'^2)rA+ABB'=0,
 \ee
 which can be integrated to give
\be
A=\displaystyle{\tilde{K_1} \over \left(1-\displaystyle{rB' \over
B} \right)},
 \ee
  $\tilde{K_1}$ being a constant of
integration. Using (15) and (35) we get after some rearrangement,
 \ba \A \A D^2 \left \{
2BB''r^3-5B'^2r^3-3rB^2+8r^2BB' \right \} \nonu
\A \A +DD' \left(2B B' r^3-2 r^2 B^2 \right)+2rB^3B''-3 r B'^2
B^2+4 B^3B'=0. \ea Using  transformation (24), we get \be (2
\alpha_{zz}-3{\alpha_z}^2+6\alpha_z-3)+{ \beta_z \over
\beta}(\alpha_z-1) +{1 \over \beta} e^{(2\alpha -2z)}(2
\alpha_{zz}-{\alpha_z}^2+2\alpha_z)=0, \ee where $\alpha_z$ is
defined above. The solution of (37) is given by, \be
\beta=D^2={e^{3(\alpha-z)} \over (1-\alpha_z)^2} \left \{
\tilde{K_2}-\alpha_z (\alpha_z-2)e^{(z-\alpha)} \right \}, \ee
where $\tilde{K_2}$ is another constant of integration. Hence the
general solution, having spherical symmetry, of the system
(15)-(19) is given by,
 \ba
 A \A =\A
\displaystyle{\tilde{K_1} \over \left(1-\displaystyle{rB' \over B}
\right)}, \nonu
 D^2 \A =\A \displaystyle{1 \over \left(1-\displaystyle{r B' \over B} \right)^2} {\left( B \over r\right)^3}
\left \{ \tilde{K_2}-{r B' \over B} \left( {r B' \over B} -2
\right) {r \over B} \right \}.
\ea
 It is clear from (39)
that as $r\rightarrow \infty$, $\tilde{K_1}=1$. Since $B$ is an
arbitrary function, we are going to consider the following case.

 If we take the arbitrary function $B$ in the form
\be
B= \displaystyle{1 \over \left(1+\displaystyle{c_1 \over r^2}
\right)^{1/2}},
 \ee
  where $c_1$ is a constant, then substituting
(40) in (39), we get

\ba
A \A =\A \left(1+\displaystyle{c_1 \over r^2} \right) \nonu
 D^2 \A =\A { \tilde{K_2}(r^2+c_1)^{3/2}+ c_1(2r^2+c_1) \over r^4(r^2+c_1) }.
 \ea
It is of interest to note that a solution  similar to that  given
by (40) and (41)  was obtained before by  one of the authors
\cite{W5} in case of the generalized field theory.  From now on,
we will take into account Lorentz signature for the solution (40)
and (41). Using (14) the line-element, in which Lorentz signature
is inserted can be written in the form, \be dS^2=-{(B^2-D^2 r^2)
\over A^2 B^2} dt^2-2{D r \over A B^2} dr dt + {1 \over B^2}
\left( dr^2+r^2 d\Omega^2 \right). \ee We can eliminate the cross
term from (42), by using the coordinate transformation (30), \be d
\tilde{T}=dt-{D r A \over (B^2- D^2 r^2)}dr, \quad \eta={(B^2-D^2
r^2) \over A^2 B^2}, \quad and \quad \tilde{R}={ r \over B}, \ee
then
 we finally get
\be
dS^2=-\eta d \tilde{T}^2 + {1 \over \eta} d
\tilde{R}^2+\tilde{R}^2 d\Omega^2.
 \ee
 Using (40) and (41), we get
\be
\eta(\tilde{R})=\left(1-{\tilde{K_2} \over \tilde{R}} -{4c_1 \over
\tilde{R}^2}+{2{c_1}^2 \over \tilde{R}^4} \right), \ee which is
different from Schwarzschild metric.

Tetrad (13) in {\it Case II} has been subject to two steps of
coordinate transformations from $(t, r, \theta, \phi)$ to
$(\tilde{T}, \tilde{R}, \theta, \phi)$. We now apply a further
transformation from $(\tilde{T}, \tilde{R}, \theta, \phi)$ to the
Cartesian coordinate $(\tilde{T},X^a)$ with $a=1,2,$ and 3. The
tetrad can be shown to  have the form

 \ba
  {\blz}^0 \A=\A \displaystyle{(1-\displaystyle{c_1 \over r^2}) \over (1-\displaystyle{k_2\over r}
  -\displaystyle{4c_1 \over r^2}+\displaystyle{2c_1^2\over r^4})} , \nonu
{\blz}^\alpha \A=\A -\displaystyle{x^\alpha \over r^{3/2}}{
\left(k_2+\displaystyle{2c_1 \over r}-\displaystyle{c_1^2 \over
r^3}\right)} , \nonu
{\al}^0 \A = \A -i\displaystyle{x^\alpha \over
r^{3/2}}{{\left(k_2+\displaystyle{2c_1 \over
r}-\displaystyle{c_1^2 \over r^3}\right)} \over
\left(1-\displaystyle{k_2 \over r} -\displaystyle{4c_1 \over
r^2}+\displaystyle{2c_1 \over r^4}\right)} , \nonu
{\al}^\alpha \A = \A -i\displaystyle{(c_1 x^\alpha x^a -r^4) \over
r^4}.
 \ea
Tetrad (46) in its Cartesian form will be used in the next section
in the calculation of the energy for the
 positive-definite case. It is of interest to note that the
 Schwarzschild metric can be obtained in this case upon taking
 $c_1=0$.

\newsection{Calculations of Energy}
In this section we are going to calculate the energy associated
with the two solutions obtained in the previous section using the
superpotential method given by Mikhail \cite{MWHL}.

The superpotential of M\o ller  theory \cite {M} is given by
Mikhail et al. \cite{MWHL} as \be {{\cal U}_{\mu}}^{ \nu \lambda}
\stackrel{\rm def.}{=}{\sqrt{-g} \over 2 \kappa} {P_{\chi \rho
\sigma}}^{\tau \nu \lambda} \left[C^\rho g^{\sigma \chi} g_{\mu
\tau}-\lambda g_{\mu \tau} \gamma^{\chi \rho
\sigma}-(1-2\lambda)g_{\mu \tau} \gamma^{\sigma \rho \tau}
\right], \ee where ${P_{\chi \rho \sigma}}^{\tau \nu \lambda}$ is
\be {P_{\chi \rho \sigma}}^{\tau \nu \lambda} \stackrel{\rm
def.}{=}{\delta_\chi}^\tau {g_{\rho \sigma}}^{\nu
\lambda}+{\delta_\rho}^\tau {g_{\sigma \chi }}^{\nu
\lambda}-{\delta_\sigma}^\tau {g_{\chi \rho}}^{\nu \lambda}, \ee
with ${g_{\rho \sigma}}^{\nu \lambda}$ being a tensor defined by
\be {g_{\rho \sigma}}^{\nu \lambda} \stackrel{\rm
def.}{=}{\delta_\rho}^\nu
{\delta_\sigma}^\lambda-{\delta_\sigma}^\nu {\delta_\rho}^\lambda,
\ee and $ \gamma^{\chi \rho \sigma}$ is the contorsion defined by
\be
 \gamma_{\mu \nu \rho} \stackrel{\rm def.}{=}{\il}_\mu {\il}_{\nu;\rho},
 \ee
 and $\lambda$ is a free dimensionless parameter.
 The energy is given by the surface integral (cf.\cite{MWHL})
\be E \stackrel{\rm def.}{=}lim_{\rho\rightarrow \infty}
\int_{\rho=constant} {{\cal U}_0}^{0 \alpha} n_{\alpha} dS, \ee
where $n_\alpha$ is the unit 3-vector normal to the surface
element $dS$. The superpotential  associated with the solution in
the case of indefinite metric (33) is given by \be {{\cal
U}_{0}}^{0 \alpha}={X^\alpha \over 4 \pi r^3} \left[2m-r^2B'
\right]. \ee Substituting (52) into (51), we obtain \be
E(r)=2m-B'r^2.
 \ee
  It is clear from (53) that the energy associated
with (33) is dependent on the arbitrary function $B(r)$ which does
not appear in the line-element (31) but appears in the tetrad
(33).

For the case of positive-definite solution (46), the
superpotential is given by
 \be
 {{\cal U}_{0}}^{0\alpha}= {X^\alpha \over 4 \pi r^3}
\left[k_2+{3c_1 \over r} \right]. \ee
 Substituting (54) into (51), we obtain
 \be
E(r)=k_2+{3c_1 \over r}.
\ee
\newsection{Comparison and Discussion}
%{\large {\bf Conclusion and discussion}}

In the present work we have studied the effect of signature change
on the solutions of the field equations of NGR. As  is pointed out
in \cite{KS},  NGR is a gravitational theory formulated using the
AP-space. Any AP-structure is defined completely, in 4-dimensions,
by a tetrad vector field subject to the AP-condition as stated  in
section 2. The importance of examining gravity theory written in
this geometry is that tetrads play an important role in many
aspects, e.g. they are used as  fundamental variables in the
attempts  to quantize gravity \cite{RH}. So, one expects to get,
using such theories, more information about gravity than those
obtained from general relativity. This is one of the goals of the
present study. In order to compare the results obtained with the
corresponding results of general relativity, we have chosen the
case having spherical symmetry for application. Also we have
solved the field equations of NGR in free space, i.e. $T^{\mu
\nu}=0$.

As mentioned in the introduction, there are two philosophies
behind the idea of signature change. The first assumes that the
change of signature, from a positive -definite 4-dimensions to an
indefinite (3+1)-dimensions,  occurred at a certain epoch when the
universe was younger, i.e. it occurred on a hypersurface of the
space-time (cf. \cite{ES,TM,E,JH}).
 The second philosophy assumes that the
universe  has a positive-definite metric, always and everywhere,
with 4-dimensions while our measurements and observations are
carried out in  (3+1)-dimensions with an indefinite metric
\cite{W}. Consequently, there is no need to impose the Lorentz
signature on the metric (from which we formulate the differential
equations) before solving these equations
 but this signature is to be imposed on the
solutions of these equations, i.e. just before matching the
results of the theory with measurements  and observations.
 Although the first approach has some advantages, yet there is a
singularity at the change surface \cite{E}. Using the second
philosophy, one can overcome this difficulty.

Calculations in the present work have been done following the
second philosophy. To study the effect of signature change, we
have solved the field equations of NGR in  two different cases. In
the first case we have imposed the Lorentz signature on the metric
(imaginary tetrad) before formulating the differential equations
as usually done in the scheme of GR. In the second case we have
formulated and solved the differential field equations using a
positive-definite metric (real tetrad), then we imposed the
Lorentz signature on the solution obtained in order to compare the
results with the well known physics of GR. The same procedure has
been previously used in the case of general relativity \cite{W}.
The results obtained are similar to those obtained from general
relativity. Table I summarizes these results and compares them
with those  previously obtained by Hayashi and
Shirafuji \cite{KS}.\\
%\newpage
\begin {center}
\begin {table} [h] \caption{ Comparison between the present solutions and Hayashi-Shirafuji Solution}
\begin {tabular} {|c|c|c|c|} \hline
  & Case I & Case II & Hayashi-Shirafuji solution\\ \hline
Tetrad ${\bl}_\mu$ & imaginary, non-diagonal & real, non-diagonal
& imaginary, diagonal\\ \hline Metric & indefinite & +ve definite
& indefinite
\\ \hline Number of solution & one & many & one \\ \hline
 Schwarzschild solution & yes & yes & yes \\ \hline Second-order skew
 tensors & some identically  vanishing & all non-vanishing & all vanishing\\ \hline
 \end {tabular}
 \end {table}
\end {center}

From  table I it is clear that, although the Schwarzschild metric
is obtained in the three cases, these cases are associated with
different sets of  second-order skew tensors. As it is clear, the
diagonal tetrad used by Hayashi and  Shirafuji \cite{KS} does not
produce any skew tensors of the second order. So, if some physics
is to be attributed
 to  these skew tensors, then the physical contents of the three cases, given in
 table II
 are quite different. The role of these second-order skew tensors is totally
obscured in the case of Riemannian geometry, since such tensors
are not defined in this geometry. For this reason general
relativity was written in the AP-geometry when examining the
effect of signature change \cite{W}. It can be shown that the
non-vanishing second-order skew tensors depend on the function
$D(r)$, whose vanishing will give rise to the vanishing of those
skew tensors. The role of such tensors can be clarified if  field
theory dealing with other interactions together with gravity. Such
a situation was achieved in the case of the generalized field
theory \cite{MW}. Referring to the discussion given in
\cite{W,W5,WB},
 it is shown that the non-vanishing skew tensors are related to
the electromagnetic field.

On the other hand, in the solution given by (40) and (41) there
are two different constants of integration $c_1$ and
${\tilde{K_2}}$. If we take $c_1=0$, the solution will give rise
to the Schwarzschild metric. Then, we can identify ${\tilde{K_2}}$
with the geometric mass of the source of the gravitational field.
While, if $c_1\neq 0$, then if we evaluate the metric (49) far
from the source of the field, i.e. neglecting the term
$O\displaystyle{1 \over (\tilde{R}^4)}$, the metric will be
similar to Reissner-Nordstr$\ddot{\rm o}$m metric. This may
indicate that the constant $c_1$ can be related to the electric
charge of the source.

To investigate the structure of the two solutions (22), (27) and
(40), (41) we make a physical application by calculating the
energy associated with these two solutions using the
superpotential potential method given by Mikhail et al.
\cite{MWHL}. We transform the tetrad which gives the Schwarzschild
solution to the Cartesian form to calculate the energy. As  is
clear from (53)  the energy depends on the arbitrary function
$B(r)$. If the asymptotic behavior is of $O(1/r)$ then the form of
the energy becomes
 $E=2m+ some\ cont.$, which in general is different from the general relativity in
spite of the fact that the associated form of the line-element of
these solution is the Schwarzschild. This is because  the
asymptotic form of this tetrad behaves  like $O(1/\sqrt{r})$
\cite{SN}. We also calculate the energy associated with the second
solution after transforming it to the Cartesian form using the
superpotential method. As is clear from (55)  we keep only up to
$O(1/r)$. This may indicate that the structure of the two
solutions  different as we analyze above using the skew tensors
argument.

\begin {center}
\begin {table} [h] \caption{ Comparison between the Energy of Different Solutions}
\renewcommand{\arraystretch}{2}
\begin {tabular}{|c|c|} \hline
Solution   & Superpotential Method  \\ \hline Hayashi-Shirafuji
diagonal solution & $E=m$\\ \hline Present work,  first solution
Case I& $E=2m+$ some\ cont. if $B(r)\sim O(1/r)$\\ \hline
 Present work, second solution Case II& $E(r)=k_2+\displaystyle{3c_1 \over r}$ \\\hline
 \end {tabular}
 \renewcommand{\arraystretch}{1}
 \end {table}
\end {center}
It is clear from table II that the energy of Case II depends on
$r$ which is similar to Reissner-Nordstr$\ddot{\rm o}$m energy.

\end{document}